\documentclass[a4paper]{jpconf}
\usepackage{color}
\usepackage{graphicx}
\begin{document}

\title{Strangeon and Strangeon Star}

\author{Xiaoyu Lai,$^1$ Renxin Xu$^{2,3}$}

\address{$^1$ School of Physics and Engineering, Hubei University of Education, Wuhan 430205, China\\
$^2$ School of Physics, Peking University, Beijing 100871, China\\
$^3$ Kavli Institute for Astronomy and Astrophysics at Peking University, Beijing 100871, China}

\ead{laixy@pku.edu.cn, r.x.xu@pku.edu.cn}

\begin{abstract}
The nature of pulsar-like compact stars is essentially a central question of the fundamental strong interaction (explained in quantum chromo-dynamics) at low energy scale, the solution of which still remains a challenge though tremendous efforts have been tried.
This kind of compact objects could actually be strange {\em quark} stars if strange quark matter in bulk may constitute the true ground state of the strong-interaction matter rather than $^{56}$Fe (the so-called Witten's conjecture).
From astrophysical points of view, however, it is proposed that strange cluster matter could be absolutely stable and thus those compact stars could be strange {\em cluster} stars in fact.
This proposal could be regarded as a \underline{general Witten's conjecture}: strange matter in bulk could be absolutely stable, in which quarks are either free (for strange quark matter) or localized (for strange cluster matter).
Strange cluster with three-light-flavor symmetry is renamed {\em strangeon}, being coined by combining ``strange nucleon'' for the sake of simplicity.
A strangeon star can then be thought as a 3-flavored gigantic nucleus, and strangeons are its constituent as an analogy of nucleons which are the constituent of a normal (micro) nucleus.
The observational consequences of strangeon stars show that different manifestations of pulsar-like compact stars could be understood in the regime of strangeon stars, and we are expecting more evidence for strangeon star by advanced facilities (e.g., FAST, SKA, and eXTP).
\end{abstract}

\section{Introduction: {\it What is a strangeon}?}

The baryonic matter in the Universe is well described by the Standard Model of particle physics.
Six flavors of quarks are fundamental Fermions that constitute baryonic matter, among which up ($u$), down ($d$) and strange ($s$) quarks are light flavors (the other heavy flavors: $c, t, b$).
Atomic nucleus, the most familiar baryonic matter to us, is composed of nucleons which is further composed of only two favors of constituent quarks, i.e. $u$ and $d$ quarks.
In other words, the nucleon is the constituent of a 2-flavored micro-nucleus.
Similarly, the ``strangeon'' we are explaining in this contribution is the constituent of a 3-flavored macro-nucleus, and is simply an abbreviation of ``strange nucleon'' (with strangeness degree of freedom there).

The existence of 3-flavored nuclei might be superficially understood by analysis of energy scale.
Quarks are often described by mass parameters that are measured indirectly since they are confined inside hadrons rather than free particles, and the mass parameters are the key parameters to make a judgment on the quark-flavor degrees of freedom at a certain energy scale.
The current mass of both $u$ and $d$ quarks are only a few MeV while that of $s$ quark is around 100 MeV.
For nuclei or nuclear matter with baryon density $n_0\simeq 0.16 \rm\ fm^{-3}$, the separation between quarks is $\Delta x\sim 0.5$ fm, and the energy scale is then in the order of $E_{\rm nucl}\sim 400$ MeV according to Heisenberg's relation $\Delta x \cdot pc \sim \hbar c \simeq 200\ \rm MeV\cdot fm$. All of the heavy flavors of quarks have masses larger than 1 GeV, and would then not participate in dense matter with density $n\sim n_0$ as valence quarks.
One could thus naively expect that 3-flavored nuclei could exist in case of density around $n_0$ because $u$ and $d$ quarks are the lightest and $E_{\rm nucl}$ is much larger than the mass difference between $s$ and $u$/$d$ quarks.
%
%
However, it is a fact that the valence strangeness degree of freedom is absolutely missing in stable nuclei with sizes $\sim$ fm.
On one hand, the mass of $s$ quark is larger than that of $u$/$d$ quarks, so $s$ quarks would convert to $u$ and $d$ quarks via weak interaction.
But, on the other hand, the light-flavor ($u$, $d$ and $s$) symmetry might be restored if the flavor symmetry leads to lowering the energy (as for an example of nuclear analogue, the symmetry energy), especially in case that the strong-interaction matter is very big.
We are trying to explain in this proceeding that 3-flavored matter could be manifested in the form of ``gigantic nucleus'', and that ``strangeon'' is the constituent of such a 3-flavored ``nucleus''.

To demonstrate the difference between two- and three-flavored nuclei, we could first look at the structure of an atom.
Before the era of Ernest Rutherford who was the father of nuclear physics, the structure of atoms was described in J. J. Thomson's model, which states that an atom is composed of positively charged medium (occupying the whole atomic volume) as well as electrons embedded.
In 1911, E. Rutherford proposed that positively charged nucleus, carrying almost all atomic mass, is actually a tiny part of an atom while electrons are not inside but around the nucleus, according to the experimental results of deflecting $\alpha$-particles passing through a thin gold foil~\cite{Rutherford1911}.
In the Standard Model of particle physics, ordinary nuclei, composed of nucleons, have only two flavors of quarks ($u$ and $d$) in fact.
Conceptually, why is our baryonic matter symmetric with two flavors rather than three?

A philosophical explanation could be simple: micro-nuclei are too small to have three-flavor symmetry, but bigger is different~\cite{XG2016}.
The electric charges of $u$ and $d$ quarks are $+2/3$ and $-1/3$ respectively, so two-flavor symmetric strong-interacting matter should be positively charged, and electrons are needed to maintain electric neutrality.
The probability of electrons inside a micro-nucleus is negligible because the nucleus radius ($\sim$ 1 fm) is much smaller than the Compton wavelength of electrons, $\lambda_{\rm e} \sim 10^3$ fm.
Therefore, electrons contribute negligible energy there, and two-flavored micro-nuclei would be energetically economic.
However, for a gigantic nucleus with size approximating to or even larger than $\lambda_{\rm e}$ (corresponding to number of nucleons $A>10^9$), it might be three-flavor symmetric.
If the two-flavor symmetry still stays in macro-nucleus, it would become a huge Thomson atom with electrons inside it.
Electron could then contribute a significantly high Fermi energy to the total energy of the system, being $E_{\rm F}\sim 100$ MeV.
The situation becomes different if strangeness (i.e., three-flavor symmetry) is included: no electrons exist if the matter is composed of equal numbers of $u$, $d$ and $s$ quarks!
In this case, the three-flavor symmetry, an analogy of two-flavor symmetry in micro-nucleus, may result in a ground state of gigantic nuclei.
Strangeon, the constituent of a gigantic nucleus, is also an analogy of nucleon which is the constituent of a micro-nucleus.
Certainly micro-physical calculations (e.g., an extended version of relativistic mean field theory applicable to strangeon matter) are really necessary and welcome in order to have a solid foundation of the starngeon conjecture.

It is worth noting that, rational thinking about stable strangeness datas back to 1970s and bulk strange object is speculated to be the absolutely stable ground state of strong-interacting matter, which is known as Witten's conjecture~\cite{Witten1984}.
Although the conjecture was proposed based on the matter composed of almost free quarks,  we can make an extension that it still reasonably holds no matter whether quarks are free or localized.

In some of our previous papers, we use a word, ``strange quark-cluster''~\cite{Xu2003}, which has actually the same meaning as strangeon.
This paper is organized as following.
The gravity-compressed dense matter is introduced in Section 2 in order to make sense of realistic ``gigantic nucleus'' in astrophysics.
The observational consequences of strangeon stars, including the surface and global properties, are discussed in Section 3.
A brief summary is given in Section 4.

\section{Compact stars with strangeness}

``Gigantic nucleus'' is not a new concept.
In the year 1932, L. Landau published a paper in which an idea of gigantic nucleus was presented~\cite{Landau1932}: ``{\it We expect that this must occur when the density of matter becomes so great that atomic nuclei come in close contact, forming one gigantic nuclues.}''
Although the main motivation for Landau to write this paper was to explain the origin of stellar energy (and this explanation is certainly wrong), he recognized for the first time a kind of compact stellar object at about nuclear density, which is surely necessary for us to understand various extreme phenomena through astronomical observations.
Two years later, in 1934, W. Baade and F. Zwicky conceived the idea that forming neutron stars could be the energy source of supernova rather than of stars~\cite{BZ1934}.
Neutron stars theoretically predicted were finally discovered when Hewish and Bell detected radio pulsars in 1967~\cite{Hewish1968}.
The gigantic nucleus proposed by Landau was then identified by astronomers to be ``neutron stars'', as neutrons and protons at that time were believed to be elementary particles and gigantic nucleus should be neutron-rich.
%
%
After physicists realized that the more fundamental baryonic particles are quarks, some models were built about compact stars including quark degree of freedom, either in the core of neutron star (i.e. mixed or hybrid stars) or as the whole star (quark stars).

One may speculate the nature of dense matter inside pulsar-like compact stars as following.
After core-collapsing of an evolved massive star, the supernova-produced rump left behind where normal nuclei are intensely compressed by gravity to form ``compressed baryonic matter'' which could manifest the behaviors of pulsar-like compact stars.
The average density of compressed baryonic matter, or compact stars, should be supra-nuclear density (a few nuclear saturation densities) due to gravitational force.
Dense matter may change from a hadronic phase to a deconfined phase as baryon density increases, but a very serious problem is: can the density of realistic compact stars be high/low enough for quarks to become deconfined/confined?

We argue that gravitationally compressed baryonic matter might be in a state of strangeon matter by starting from deconfined quark state with the inclusion of strong interaction between them.
With the Dyson-Schwinger-Equation approach to the non-perturbative quantum chromodynamics (QCD), one would estimate the strong coupling constant $\alpha_{\rm s}$~\cite{Fischer2006},
\begin{equation}
\alpha_{\rm s}=\frac{\alpha_{\rm s}(0)}{\ln (e+a_1x^{a_2}+b_1x^{b_2})}, \label{alpha}
\end{equation}
where $\alpha_{\rm s}(0)=2.972$, $a_1=5.292$ GeV$^{-2a_2}$, $a_2=2.324$, $b_1=0.034$ GeV$^{-2b_2}$, $b_2=3.169$, $x=p^2$ with $p$ the typical momentum in GeV.
In compact stars with quark number density $n\sim 3 n_0$, $p^2\simeq 0.16$ GeV$^2$, so the coupling parameter $\alpha_{\rm s}$ would be even larger than 2.
This means that a weakly coupling (i.e., non-perturbative) treatment could be inadequate for realistic dense matter in compact stars.
It is also worth noting that the dimensionless electromagnetic coupling constant is about $1/137<0.01$, which makes quantum electromagnetic dynamics (QED) tractable in a perturbative way.
That is to say, a weakly coupling strength comparable with that of QED is possible only if the baryon number density $n_{\rm B}>10^{123}n_0$,  being unbelievable and unrealistically high.
From this point of view, although some efforts have been made to understand the state of pulsar-like compact stars in the framework of conventional quark stars, including the MIT bag model with almost free quarks~\cite{Alcock1986} and the color-superconductivity state model~\cite{Alford2008}, realistic stellar densities cannot be high enough to justify the use of perturbative QCD which most of compact star models rely on.
The strong coupling between quarks may naturally render quarks grouped in quark-clusters, and each quark-cluster/strangeon is composed of several quarks condensating in position space rather than in momentum space~\cite{Xu2003}.

What's new about the concept of gigantic nucleus? In a word, the answer is that: Landau's gigantic nucleus is 2-flavored, while ours 3-flavored.
The compact star composed of quark-clusters could also be regarded as a gigantic nucleus, then why should the gigantic nucleus be necessarily 3-flavored?
A 3-flavored gigantic nucleus consists $u$, $d$ and $s$ quarks grouped in strangeons, while a 2-flavored micro-nucleus consists $u$ and $d$ quarks grouped in nucleons.
The only difference between gravitationally compressed baryonic matter and micro-nucleus could be a simple change from non-strange to strange: ``2''$\rightarrow$``3''.
Certainly, the state of compressed baryonic matter in compact stars is essentially a non-perturbative QCD problem and is difficult to answer from first principles.
Nevertheless, strangeon matter in bulk may constitutes the true ground state of strong-interacting matter rather than $^{56}$Fe, and this could be seen as a generalized Witten's conjecture, while the traditional Witten's conjecture focuses on the matter composed of almost free $u$, $d$ and $s$ quarks.
Consequently, we suggest that a gigantic nucleus is actually condensed matter of strangeons which is 3-flavored.

\begin{figure}[h]
\begin{center}

 \includegraphics[width=5 in]{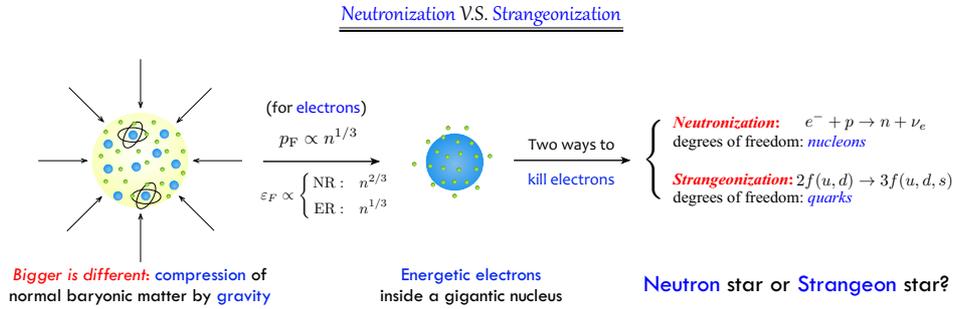}

\end{center}
\caption{
 Neutronization and strangeonization are two competing ways to cancel energetic electrons~\cite{XG2016}.
\label{fig1}}
\end{figure}

The weak equilibrium among $u$, $d$ and $s$ quarks is possible, instead of simply that between $u$ and $d$ quarks.
At the late stage of stellar evolution, normal baryonic matter is intensely compressed by gravity in the core of massive star during supernova.
The Fermi energy of electron is significant in compressed baryonic matter, and it is very essential to cancel the electrons by weak interaction in order to make a lower energy state.
There are two ways to kill electrons as shown in Fig.~\ref{fig1}: one is via the conventional {\it neutronization}, $e^-+p\rightarrow n+\nu_e$, where the fundamental degree of freedom could be nucleons; the other is through the unconventional {\it strangeonization}, where the degrees of freedom are quarks.
While neutronization works for removing electrons, strangeonization has both the advantages of minimizing the electron's contribution of kinetic energy and maximizing the flavor number, with the latter perhaps related to the flavor symmetry of strong-interaction matter.
These two ways to kill electrons are relevant to the nature of pulsars, corresponding to neutron star and strangeon star, respectively, as summarized in Fig~\ref{fig1}.

In a word, strangeon matter is conjectured to be the state of matter inside compact stars, where strangeons form due to both the strong and weak interactions and become the dominant components inside those stars.
Then, what could be a realistic strangeon? We know that $\Lambda$ particle (with structure of \{$uds$\}) possesses light-flavor symmetry, and one may think that a kind of strangeons would be $\Lambda$-like.
However, the interaction between $\Lambda$'s could be attractive, which would render more quarks grouped together.
Motivated by recent QCD simulations of the $H$-dibaryons (with structure $uuddss$), a possible kind of strangeons, $H$-clusters, is proposed~\cite{LGX2013}, and it is found that H-cluster matter could be more stable than nuclear matter when the in-medium stiffening effect is taken into account.
Actually we are not certain about the number $N_{\rm q}$ of valence quarks inside a strangeon, but one could speculate that $N_{\rm q}$ = 6, 9, 12, and even 18.
Moreover, because of their classical behavior (as a large mass may result in a small quantum wavelength), strangeons that exist in compact stars could locate in periodic lattices (i.e., in a solid state) when temperature becomes sufficiently low.
Pulsars are in the low temperature limit, and they could then be solid strangeon stars.

\begin{figure}[h]
\begin{center}

 \includegraphics[width=4 in]{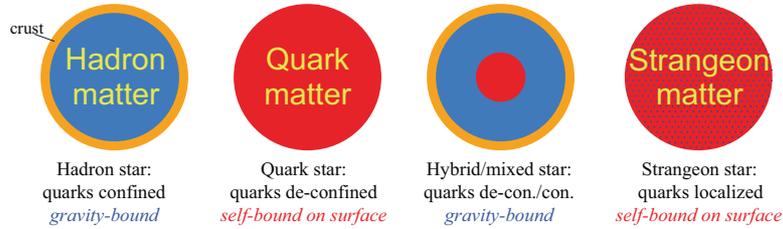}

\end{center}
\caption{
Different models of pulsar's nature. Hadron star and hybrid/mixed star belong to conventional neutron stars. Strangeness plays an important role for strange quark star and strangeon star as a result of three-light-flavor ($u$, $d$ and $s$) symmetry.
\label{fig2}}
\end{figure}

Different models of pulsar inner structure are summarized in Fig.~\ref{fig2}. The conventional neutron stars (hadron star and hybrid/mixed star) are gravity-bound, while strange stars (strange quark stars and strangeon stars) are self-bound on surface by strong force.
Let's explain in details, with attention paid to the differences among these models.
In the neutron star picture, the inner and outer cores and the crust keep chemical equilibrium at each boundary, so neutron star is bound by gravity.
The core should have a boundary and is in equilibrium with the ordinary matter because the star has a surface composed of ordinary matter.
There is, however, no clear observational evidence for a neutron star's surface.
Being similar to traditional strange quark stars, strangeon stars have almost the same composition from the center to the surface, and the strangeon matter surface could be natural for understanding some different observations, as shown in the next section.
Strangon stars are self-bound by the residual interaction between strangeons, while strange quark stars by bag-like strong interaction.
%

\section{Observational consequences expected in a strangeon star model}

Observations of compact stars, depending upon surface and global properties, could provide hints for the state of compressed baryonic matter, as discussed in the following subsections.

\subsection{Surface properties}

{\it (1) Drifting sub-pulses.}
Although pulsar-like stars have many different manifestations, they are populated by radio pulsars, and it seems that most of radio sub-pulses are drifting.
Among the magnetosphere-dynamics models for pulsar radiative process, the Ruderman-Sutherland~\cite{RS1975} model is a virtue not shared by others, and the drifting sub-pulses were first explained.
The RS model suggests strong binding of particles on pulsar polar caps, but the calculated binding energy in neutron star models could not be so high.
This problem could be naturally solved in bare strangeon star scenario due to the strong self-bound of strangeons on surface.

The magnetospheric activity of bare strangeon star was investigated in quantitative details~\cite{YX2011}.
Since strangeons on the surface are confined by strong color interaction, the binding energy of strangeons can be even considered as infinity compared to electromagnetic interaction.
As for electrons on the surface, calculations have shown that the huge potential barrier built by the electric field in the vacuum gap above the polar cap can usually prevent electrons from streaming into the magnetosphere.
Therefore, in the strangeon star model, both positively and negatively charged particles on the surface are usually bound strongly enough to form a vacuum gap above its polar cap, and the drifting (even bi-drifting) sub-pulses can be understood naturally~\cite{Xu1999}.
Certainly, more researches, both theoretical and observational, on this topic are welcome since this subpulse drifting phenomenon could hold the key to understanding the magnetospheric activity as well as the surface material of pulsar.

{\it (2) Clean fireball for SNE/GRBs.} It is still an unsolved problem to simulate supernovae successfully in the neutrino-driven explosion models of neutron stars.
Nevertheless, in the strangeon star scenario, the bare strangeon matter surface could be essential for successful explosions of both core and accretion-induced collapses~\cite{Xu2005}.
A nascent strange star born in the center of GRB or supernova would create a thermal fireball due to its ultrahigh surface temperature~\cite{Haensel1991}, and the photon luminosity is not constrained by the Eddington limit since the surface of strange star could be bare and chromatically confined~\cite{Ouyed2005,Paczynski2005}.
Therefore, in this photon-driven scenario the strong radiation pressure caused by thermal emission from strangeon star might play an important role in promoting core-collapse supernovae~\cite{CYX2007} and even in explaining long-lived plateau of light curves of GRB afterglow~\cite{DLX2011}.

\begin{figure}[h]
\begin{center}

 \includegraphics[width=8 cm]{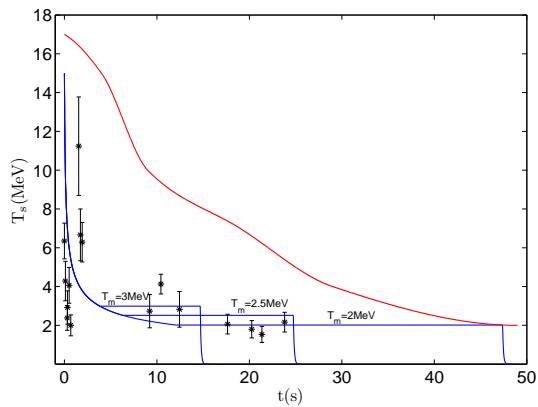}

\end{center}
\caption{
This is the $T$(temperature)-$t$(time) relation of a proto-strangeon star with mass $1.4M_{\odot}$ and radius $10$ km~\cite{Yuan2017}. We take logarithm of age $t$ in order to see the details. The blue lines are $T-t$ curves for strangeon star with different parameters of melting temperature $T_{\rm m}$, and the upper red line is for a normal proto-neutron star taken from~\cite{Pons1999}. The 15 black dots with error bars are 15 neutrino events, of which the neutrino energy ($E_\nu$) has been represented by proto-strangeon star's temperature ($T$) by a relation of $E_{\nu}=3.15 T$. It is evident that temperature $T$ of a proto-strangeon star drops down rapidly in millisecond timescale, and then the star will keep homothermal during a liquid-solid phase transition, as the straight lines indicate. After the phase transition, solid strangeon stars cool down more drastically than ever before. In this case, consequently, the emission intensity decreases quickly, and hence leading to a cut-off of this neutrino burst. On the contrast, normal neutron stars cool down smoothly all the time, and have no interceptive indication during supernova neutrino burst. It indicates that strangeon star model with melting temperature $\sim 1$ MeV coincides with the burst events, either the duration or the energy of neutrino radiation, detected from SN1987A.
\label{fig3}}
\end{figure}

The neutrino burst observed during supernova 1987A could also be understood in the regime of strangeon star~\cite{Yuan2017}.
Huge internal energy is stored in a newborn strangeon star after collapse, and then the energy is released by photons and neutrinos, but being dominated initially by neutrino radiation.
A liquid-solid phase transition at temperature $\sim 1$ MeV may occur only a few ten-seconds after core-collapse, and the thermal evolution and the neutrino emission of a strangeon star could then be modeled, as illustrated in Fig.~\ref{fig3}.
%

{\it (3) Non-atomic spectral features.} The existence of atomic features in the thermal X-ray emission of neutron star atmosphere is expected in neutron star models.
Differently, a bare strangeon star does not have atomic features in the thermal X-ray emission because no atom exists on its surface.
Note that slight plasma may cover on a strangeon star surface, which is optically thin in X-ray bands (see ``{\it (5)}'' below for discussions).
The featureless spectrum predicted by strangeon star model is consistent with observations of {\it Chandra} and {\it XMM-Newton}, both of which have not detected certain atomic feature predicted in neutron star atmosphere models.
In neutron star scenario, the spectrum determined by radiative transfer in atmosphere should differ substantially from Planck-like one.

{\it (4) Absorption lines in thermal X-ray spectrum.} Although the three-light-flavor symmetry breaking is tiny (with the electric charge per baryon $\sim 10^{-5}$), the electron number density is still high and the Fermi energy of electrons is $E_{\rm F}\sim 10$ MeV, one order smaller than that in case of two-light-flavor symmetry.
Strangeon matter is bound by the strong interaction, while electrons are bound to strangeon matter by the electromagnetic interaction, so some of the electrons in the surface region of a strangeon star reside outside of strangeon matter boundary, leading to a fairly thin (thousands of femtometers thick) electron sea on the surface of strangeon matter~\cite{Alcock1986}.
The global motion of the electron sea on the magnetized surface was investigated~\cite{Xu2012}, and it is found that hydrodynamic surface fluctuations of the electron sea would be greatly affected by the magnetic field.
Some observations did show absorption lines of pulsar-like stars, and the best absorption features are detected for the central compact object (CCO) 1E 1207.4-5209, at $\sim 0.7$ keV and $\sim 1.4$ keV~\cite{Sanwal2002,Mereghetti2002,Bignami2003}.
The absorption feature of 1E 1207.4-5209 could be understood in this hydrocyclotron oscillations model~\cite{Xu2012}.
Besides the absorption lines in 1E 2107.4-5209, the detected lines around (17.5, 11.2, 7.5, 5.0) keV in the burst spectrum of SGR 1806-20 and those in other dead pulsars (e.g. radio quiet compact objects) would also be of hydrocyclotron origin~\cite{Xu2012}.
It seems that X-ray absorption feature could be natural in strangeon star model.

{\it (5) Strangeness barrier.}
The constituent quarks in a strangeon are of three flavors ({\em u}, {\em d} and {\em s}) rather than of two flavors ({\em u} and {\em d}) for normal nucleons.
Consequently, the fundamental weak interaction does play an essential role to convert normal 2-flavored matter (i.e., nucleons) to 3-flavored one (i.e., strangeons) during an accretion phase.
The weak-conversion, however, is not easy, and could be successful only after frequent collisions (order of $\gg 1$), similar to the famous $pp-$reaction with  flavor change. We would then introduce a term of {\em strangeness barrier} to describe this kind of difficulty~\cite{Xu2014}.
Because of this barrier above strangeon star surface, which separates two-flavor matter from three-flavor matter, most of the ordinary nuclei falling onto a strangeon star might bounce back.
Therefore, a strangeon star may be surrounded by a hot corona or an atmosphere, or even a crust for different accretion rates.
The strangeness barrier could help us to understand the redshifted O VIII Ly-$\alpha$ emission line (only with $z=0.009$) and the change in the blackbody radiation area of 4U 1700+24 if it is a low mass strangeon star~\cite{Xu2014}, as well as the low mass function measured for the puzzling X-ray binary.

\begin{figure}[h]
\begin{center}

 \includegraphics[width=\textwidth]{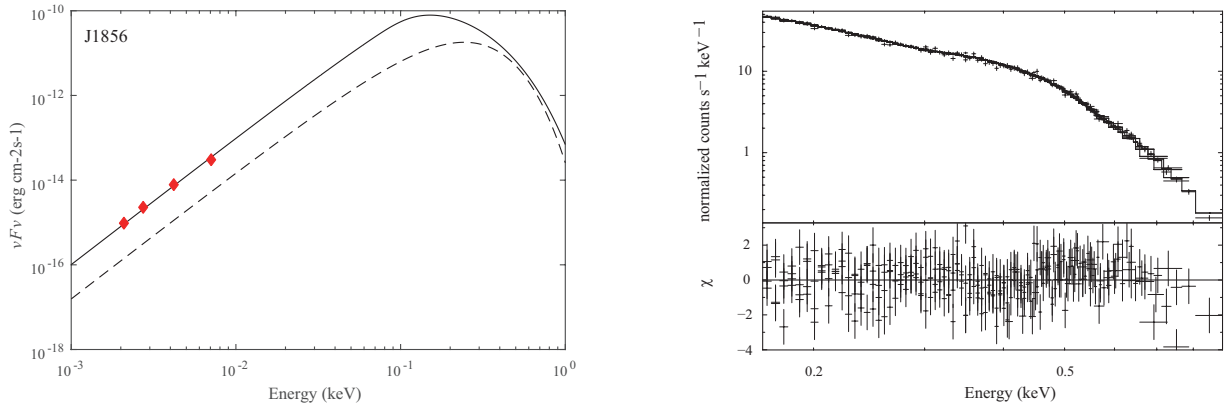}

\end{center}
\caption{
The bremsstrahlung emission (left, solid line) and X-ray data fitting (right, with $\chi^2/{\rm dof}=1.17/298$) from a strangeon star atmosphere are shown for RX J1856.5-3754~\cite{Wang2016}. By comparison, a dashed line of pure black-body model extrapolated from its X-ray spectrum is also drawn. The red diamonds are of optical observations from the {\em HST} photometry.
The extracted spectra are binned with 1000 counts per bin in each observation at least. The X-ray data are fitted by the bremsstrahlung plus a Gaussian function which indicates an absorption line around 0.21\,keV ($\sigma=0.02$\,keV).
\label{fig4}}
\end{figure}

Additionally, the strangeness barrier could also be meaningful to understand Type-I X-ray bursters as well as to constrain their masses and radii~\cite{Li15}.
Despite these, the barrier could also be necessary to solve the optical/UV excess puzzle with a strangeon star atmosphere.
X-ray dim isolated neutron stars (XDINSs) are characterized by Planckian spectra in X-ray bands, but show optical/ultraviolet (UV) excesses exptrapolated from X-ray spectra.
In the regime of strangeon stars, the ISM-accreted matter could form a plasma atmosphere due to strangeness barrier.
A radiative model of bremsstrahlung emission from the plasma atmosphere could fit well the spectra of the seven XDINSs, from optical/UV to X-ray bands, as shown in Fig.~\ref{fig4} for the famous one, RX J1856.5-3754~\cite{Wang2016}.
%
%
It is worth noting that this strangeon star atmosphere could simply be regarded as the upper layer of a normal neutron star atmosphere, but with almost homogeneous electron (or ion) temperature.
Therefore, in the strangeon star atmosphere, thermal X-rays from lower layer of normal neutron star atmosphere are prohibited, and relatively more optical/UV photons are then radiated.
Hard X-ray cut-off (i.e., without a hard tail) would also be natural in our model.
Furthermore, the optical/UV spectral deviations from a Rayleigh-Jeans distribution could come from non-uniformities of the atmosphere~\cite{Wang2017}.

\subsection{Global properties}

{\it (1) Stiff equation of state and massive strangeon star.} The equation of state of strangeon matter would be stiffer than that of nuclear matter for two main reasons: 1) strangeons should be non-relativistic particles for its large mass, and 2) there could be strong short-distance repulsion between strangeons (an analogy for the hardcore of nuclear force).
It is also worth noting that, both the problems of hyperon puzzle and quark-confinement do not exist in strangeon star model, whereas those two are still challenging in conventional neutron star and quark star models.

It has been addressed that strangeon stars could have high maximum masses ($>2M_{\odot}$) inferred by the stiff equation of state, as well as very low masses ($< 10^{-2} M_{\odot}$) as a direct consequence of self-bound surface~\cite{LX2009}.
Later observations of PSR J1614-2230 imply a mass of $1.97\pm 0.04 M_{\odot}$~\cite{Demorest2010}, indicating a stiff equation of state.
It is conventionally though that the state of dense matter softens and thus cannot result in high maximum mass if pulsars are quark stars.
However, strangeon stars could not be ruled out by massive pulsars, and the observations of pulsars with higher mass (e.g. $>2.5 M_{\odot}$) would even be a strong support to strangeon star model~\cite{LGX2013, GLX2014}.
New observation indicates that the mass of PSR J2317+1439 would even be larger than $2.04 M_\odot$ (at 90\% confidence)~\cite{Dai2017}, being expected in strangeon star model.

{\it (2) Self bound and low mass strangeon stars.} As discussed previously, similar to strange quark stars, strangeon stars are bound by strong force instead of gravity.
Self-bound strangeon stars have non-zero surface density, and their radii usually increase as masses increase, while for gravity-bound neutron stars the radii decrease as mass increase.
Gravity only play a significant role when the mass of the strangeon star approaches the maximum mass, beyond which the (repulsive) interaction between strangeons could no longer resist the gravity and the star would collapse to become a black hole.
Specially, $M\propto R^3$ for low-mass strangeon stars with negligible gravity, so the mass of strangeon stars could be very low with masses as low as $\sim 10^{-2} M_{\odot}$ ~\cite{Xu2005} (being even manifested in the form of strangeon planet~\cite{XW2003}).

{\it (3) Anisotropic pressure and extra-free energy besides powers of spin, accretion and heat.} Strangeon stars could be in a solid state at low temperature.
The local pressure could be anisotropic in elastic matter, so for solid strangeon stars the radial pressure gradient could be partially balanced by the tangential shear force.
Release of the elastic energy as well as gravitational energy would not be negligible, and may have significant astrophysical implications.

A sudden change of tangential force may result in a starquake, with release of both gravitational and elastic energy. There are actually two kinds of starquakes in the solid strangeon star model: bulk-invariable (type I) and bulk-variable (type II) starquakes, with energy release negligible for the former but significant for the latter.
Typical energy of $10^{44-47}$ erg is released during superflares of soft gamma-ry repeaters (SGRs), and a type II giant starquake could reproduce such a flare~\cite{XTY2006}.
Both types of glitches with (type II) and without (type I) X-ray enhancement could be naturally understood in the starquake model of solid strangeon stars~\cite{Zhou2014}.

Significant stellar oscillations could usually accompany a starquake, especially for the Type II quakes.
The magnetospheric activity in the polar cap region of pulsars could be excited under such oscillations. The toroidal oscillation of the star propagates into the magnetosphere, which provides additional voltage due to unipolar
induction, changes Goldreich-Julian charge density from the traditional value due to rotation, and hence influences particle acceleration~\cite{Lin2015}.
The onset of radio emission after glitches/flares in SGRs or AXPs (anomalous X-ray pulsars) could be the result of oscillation-driven magnetic activities of solid strangeon stars. Within that model, transient radio signals of AXPs/SGRs may be caused by activation of the pulsar inner gap from below the radiative death line, due to an oscillation-induced voltage enhancement.

{\it (4) Rigidity.} Rigid body precesses naturally when spinning, either freely or by torque, but fluid one can hardly.
The observation of possible precession or even free precession of B1821-11~\cite{Stairs2000} and others could suggest a global solid structure of compact star.
It is suggested that the gravitational wave behaviors should be mass-dependent~\cite{Xu2006}, and no gravitational wave originated from $r$-mode instability could be detected for solid strangeon stars.
Moreover, the bumpy distortion, i.e. local mountains on star surface, can lead to asymmetry about a pulsar's rotation axis, which could be tested by gravitational wave detectors in the future.
By precision pulsar timing, either in radio or in X/$\gamma$-ray bands, we may discover more and more candidates with precession, that might provide a reliable way to test the strangeon star model.
We are looking forward to these observations and discoveries.

The discovery of the gravitational waves~\cite{Abbott2016} opens a new window for exploring the Universe.
It is worth noting that, strangeon star with rigidity is quite likely to be tested by kilo-Hz gravitational wave observations of two kinds of events at least: 1) merger of pulsar-pulsar/pulsar-black hole binaries, during which the predicted waveform and the tidal effects in inspiral depend on the equation of state of supra-nuclear matter; 2) starquake of pulsar-like compact stars, during which the induced gravitational waves of compact stars may be discovered by sensitive detectors.
Certainly, electro-magnetic diagnostics is important for both events of gravitational wave radiation, to remove the degeneracy of physical parameters.

\section{Summary}

Normal micro-nuclei with two flavors, formed initially during the Big Bang nucleosynthesis, make our Universe rich and colorful.
We conjecture, from an astrophysical point of view, that a gigantic nucleus in which three-flavor symmetry is restored could form by compression of gravity during a supernova, and its constituent are strangeons made of almost equal numbers of $u$, $d$ and $s$ quarks.
That is to say, all the pulsar-like compact stars could actually be strangeon stars.
Here we improve the original idea of gigantic nuclei presented by L. Landau over 85 years ago, with the main difference being from two-flavor to three-flavor, or from non-strangeness to strangeness.
We demonstrate that the strangeon stars could be necessary to understand different manifestations of pulsar-like compact stars, and are expecting to test this model by future observation with advanced facilities, including either ground radio telescopes Five-hundred-meter Aperture Spherical radio Telescope (FAST) and Square Kilometre Array radio telescope (SKA), or space-based high energy observatories Lightweight Asymmetry and Magnetism Probe (LAMP) and enhanced X-ray Timing and Polarimetry (eXTP).

Let's summarize the paper by a famous sentence of P. W. Anderson (1932 - ): ``The ability to reduce everything to simple fundamental laws does not imply the ability to start from those laws and reconstruct the Universe''.
As for the state of supra-nuclear matter in pulsar-like compact stars, we are still quite embarrassed because, besides the many-body problem, the fundamental law of strong interaction is still unclear in the low-energy scale.

\vspace{0.6cm}
\noindent
{\bf Acknowledgments:}
We are grateful to the members at the pulsar group of Peking University.
This work is supported by the National Natural Science Foundation of China (11673002 and U1531243) and the Strategic Priority Research Program of CAS (No. XDB23010200).


\section*{References}
\begin{thebibliography}{99}


\bibitem{Rutherford1911}
Rutherford E 1911 {\it Phil. Mag.} {\bf 21} 669

\bibitem{XG2016}
Xu R X and Guo Y J 2017, in: {\it Centennial of General Relativity: A Celebration}, p. 119-146 (arXiv:1601.05607)

\bibitem{Witten1984}
Witten E 1984 {\it Phys. Rev. D} {\bf 30}, 272

\bibitem{Xu2003}
Xu R X 2003 {\it ApJ} {\bf 596} L59

\bibitem{Landau1932}
Landau L 1932 {\it Sov. Phys.} {\bf 1} 285

\bibitem{BZ1934}
Baade W and Zwicky F 1934 {\it Phys. Rev.} {\bf 46} 76

\bibitem{Hewish1968}
Hewish A, Bell S J, Pilkington J D H, et al. 1968 {\it Nature} {\bf 217} 709

\bibitem{Fischer2006}
Fischer C S 2006 {\it J. Phys. G: Part. Nucl, Phys.} {\bf 32} R253

\bibitem{Alcock1986}
Alcock C, Farhi R and Olinto A 1986 {\it ApJ} {\bf 310} 261

\bibitem{Alford2008}
Alford M J and et al. 2008 {\it Rev. Mod. Phys.} {\bf 80} 1455

\bibitem{LGX2013}
Lai X Y, Gao C Y and Xu R X 2013 {\it MNRAS} {\bf 431} 3282

\bibitem{RS1975}
Ruderman M A and Sutherland P G 1975 {\it ApJ} {\bf 196} 51

\bibitem{YX2011}
Yu J W and Xu R X 2011 {\it MNRAS} {\bf 414} 489

\bibitem{Xu1999}
Xu R X, Qiao G J and Zhang B 1999 {\it ApJ} {\bf 522} L109

\bibitem{Xu2005}
Xu R X 2005 {\it MNRAS} {\bf 356} 359

\bibitem{Haensel1991}
Haensel P, Paczynski B and Amsterdamski P 1991 {\it ApJ} {\bf 375} 209

\bibitem{Ouyed2005}
Ouyed R, Rapp R and Vogt C 2005 {\it ApJ} {\bf 632} 1001

\bibitem{Paczynski2005}
Paczynski B and Haensel 2005 {\it MNRAS} {\bf 362} L4

\bibitem{CYX2007}
Chen A B, Yu T H and Xu R X 2007 {\it ApJ} {\bf 668} L55

\bibitem{DLX2011}
Dai S, Li X L and Xu R X 2011 {\it Science China: Physics, Mechanics \& Astronomy} {\bf 54} 1541

\bibitem{Yuan2017}
Yan M, Lu J G, Yang Z L, Lai X Y and Xu R X 2017 {\it RAA} in press

\bibitem{Pons1999}
Pons J A, Reddy S, Prakash M, Lattimer J M and Miralles J A 1999 {\it ApJ} {\bf 513} 13

\bibitem{XW2003}
Xu R X and Wu F 2003 {\it Chin. Phys. Lett.} {\bf 20} 80

\bibitem{Xu2012}
Xu R X, Bastrukov S I, Weber F, Yu J W and Molodtsova I V 2012 {\it Phys. Rev. D} {\bf 85} 023008

\bibitem{Sanwal2002}
Sanwal D, Pavlov G G, Zavlin V E and Teter M 2002 {\it ApJ} {\bf 574} 61

\bibitem{Mereghetti2002}
Mereghetti S, Luca A De, Caraveo P, Becker W, Mignani R and Bignami G F 2002 {\it ApJ} {\bf 581} 1280

\bibitem{Bignami2003}
Bignami G F, Caraveo P A, Luca A De and Mereghetti S 2003 {\it Nature} {\bf 423} 725

\bibitem{Xu2014}
Xu R X 2014 {\it RAA} {\bf 14} 617

\bibitem{Li15}
Li Z S, Qu Z J, Chen L, Guo Y J, Qu J L and Xu R X 2015 {\it ApJ} {\bf 798} 56

\bibitem{Wang2016}
Wang W Y, Lu J G, Tong H, Ge M Y, Li Z S, Men Y P and Xu R X 2017 {\it ApJ}  {\bf 837} 81 (arXiv:1603.08288)

\bibitem{Wang2017}
Wang W Y, Feng Y, Lai X Y, Lu J G, Chen X L and Xu R X 2017 submitted (arXiv:1705.03763)

\bibitem{LX2009}
Lai X Y and Xu R X 2009 {\it MNRAS} {\bf 398} L31

\bibitem{Demorest2010}
Demorest P, Pennucci T, Ransom S, Roberts M and Hessels J 2010 {\it Nature} {\bf 467} 1081

\bibitem{GLX2014}
Guo Y J, Lai X Y and Xu R X 2013 {\it Chin. Phys. C} {\bf 38} 055101

\bibitem{Dai2017}
Dai S., Smith M. C., Wang S., et al. 2017, {\it ApJ} in press (arXiv:1705.02593)

\bibitem{XTY2006}
Xu R X, Tao D J and Yang Y 2006 {\it MNRAS} {\bf 373} L85

\bibitem{Zhou2014}
Zhou E P, Lu J G, Tong H and Xu R X 2014 {\it MNRAS} {\bf 443} 2705

\bibitem{Lin2015}
Lin M X, Xu R X, and Zhang B. 2015, {\it ApJ} {\bf 799} 152

\bibitem{Stairs2000}
Stairs I H, Lyne A G and Shemar S L 2000 {\it Nature} {\bf 406} 484

\bibitem{Xu2006}
Xu R X 2006 {\it Astropart. Phys.} {\bf 25} 212

\bibitem{Abbott2016}
Abbott and et al. 2016 {\it PRL} {\bf 116} 061102

\end{thebibliography}
\end{document}